\documentclass[fleqn,usenatbib]{mnras}

\usepackage[T1]{fontenc}

\DeclareRobustCommand{\VAN}[3]{#2}
\let\VANthebibliography\thebibliography
\def\thebibliography{\DeclareRobustCommand{\VAN}[3]{##3}\VANthebibliography}

\usepackage{graphicx}	
\usepackage{amsmath}	
\usepackage{amssymb}	
\usepackage[export]{adjustbox}
\usepackage{xcolor}

\title{Growth of Aggregates with Liquid-Like Ice-Shells in Protoplanetary Discs}

\author[G. Musiolik]{
Grzegorz Musiolik\thanks{E-mail: gregormusiolik@gmail.com}\\
}

\date{Accepted on 6 July 2021}

\pubyear{2021}

\begin{document}
\label{firstpage}
\pagerange{\pageref{firstpage}--\pageref{lastpage}}
\maketitle

\begin{abstract}
During the first stages of planet formation, the collision growth of dust aggregates in protoplanetary discs (PPDs) is interrupted at the bouncing barrier. Dust aggregates coated by different species of ice turn out to be helpful to shift the bouncing barrier towards larger sizes due to their enhanced sticking properties in some cases. A rarely noticed fact is that H$_2$O- and H$_2$O-CH$_3$OH-NH$_3$-ice behave liquid-like when UV-irradiated within a PPD-similar environment. Dust aggregates coated by these ice species might be damped in collisions due to the liquid-like ice-shell which would effectively results in an increase of the sticking velocity. In this work, collisions of dust aggregates covered by the liquid-like H$_2$O-CH$_3$OH-NH$_3$-ice-shell are considered numerically. The coefficient of restitution and the sticking velocity are calculated for different thicknesses of the ice-shell. The simulation predicts that an ice-shell-thickness of few microns would be sufficient to allow the growth of cm-sized clusters in the PPD. 
\end{abstract}

\begin{keywords}
protoplanetary discs -- planets and satellites:  formation -- software: simulations 
\end{keywords}



The formation of planets begins with the collision of dust particles in a protoplanetary disc (PPD). Initially, these are assumed to have a size on the order of $\sim\mu m$ and stick together in collision events. This leads to the formation of larger and fractal-like aggregates \citep{Blum2008}. In on-going hit\&stick collisions they get increasingly compact and evolve finally to tightly bounded aggregates with a typical size on the order of a millimeter. Here, the aggregates tend mainly to bouncing than sticking in collisions due to their inertia. This effect is also called as the bouncing barrier and has been examined in detail theoretically and in experiments \citep{Guettler2010,Zsom2010,Kelling2014,Kruss2016,Kruss2017}. The bouncing barrier has to be overcome to build larger, cm-sized pebbles leading to the formation of terrestrial planets \citep{Youdin2005,Johansen2007}.

Until today, several mechanisms describing a further growth beyond the bouncing barrier have been proposed, e.g. due to magnetorotational instabilities (MRI) \citep{Balbus1991}, by collisions of the "lucky winners" \citep{Windmark2012}, due to (collision) tribocharging \citep{Jungmann2018,Steinpilz2020} or even due to dust collisions at high temperatures \citep{Bogdan2019}. A recent overview of mechanisms playing a role here is given by \citet{Wurm2021}.

The impact of water-ice on the aggregates' growth beyond the bouncing barrier was often studied in recent years as well. Water ice is not only helpful to explain different effects in a PPD like the cometary morphology as recently shown by \citet{Haack2021} but moreover it can also condensate on the surface of the dust aggregates. This shell-like structure emerges behind the water-ice snowline in a sufficient distance from the central star. In the MMSN model (Minimum Mass Solar Nebula) \citep{Weidenschilling1977,Hayashi1981},
the water-ice snowline is approximately at 2 au. At this distance, water-ice has a greater surface energy than basalt or silicates and aggregates coated with water-ice show increased sticking properties \citep{Ros2013,Gundlach2014,Musiolik2016a,Musiolik2016b}. None the less, the surface energy decreases at even farther distances from the star. The reason for this behaviour is the falling temperature and as a consequence thereof a changing crystal structure of the water-ice \citep{Gaertner2017}. This finally results in vanishing advantages or even disadvantages for the sticking of water-ice compared to materials like basalt and silicates below 200 K or 2.4 au in MMSN \citep{Musiolik2019}. Thus, this mechanism can contribute to get over the bouncing barrier at most within a narrow range of 2-2.4 au. Behind this point, aggregates are not expected to be able to grow beyond the bouncing barrier.

At this point, \citet{Tachibana2017} discover an effect which might be helpful for the growth at even larger distances from the star. The authors irradiate amorphous water-ice and a mixture of H$_2$O-CH$_3$OH-NH$_3$-ice (with a molecular ratio of 5:1:1) with UV-light at low temperatures and at 10$^{-6}$ Pa. Both ice species behave liquid-like under the irradiation at certain temperature ranges. The reason for this behaviour might be reducible to the rearrangement of hydrogen bonds within the ice mixture due to photodissociation. This explanation has to be studied in more detail in future experiments however. 

For the UV-irradiated amorphous water-ice, the authors estimate a viscosity of $4\cdot10^{7}$ Pa$\cdot$s at a temperature of 60 K. For the H$_2$O-CH$_3$OH-NH$_3$-ice mixture, the viscosity is $4-7\cdot10^{2}$ Pa$\cdot$s in the temperature range of 88-112 K. In the MMSN, such temperatures are roughly reached at the distance of 5 au (which equals the position of Jupiter in the solar system). 

Aggregates coated with a layer of the liquid-like ice species might dissipate collision energy much more efficiently and show increased sticking properties. In general, three types of collision behaviour can be distinguished depending on the viscosity of the ice-shell. The Stokes number $St$ is an important quantity in this context \citep{Gollwitzer2012}. At $St\gg 1$ one can expect low damping in collisions because the ice-shell cannot support enough drag. At $St\approx 1$ the most energy is dissipated within the shell and the coefficient of restitution (COR) reaches a minimum. The transition between both regions depends on other factors as well, e.g. the material, the impact velocity or the shell-to-core size ratio. Other quantities like the Weber number (ratio of impact energy to surface energy) are helpful to determine this boundary more precisely and are often considered in literature. For very small Stokes numbers $St\ll 1$ the collision outcome is bouncing, which might sound surprising at first. Here, the liquid-like ice-shell behaves almost like a solid on the collision time-scale of the aggregates. This e.g. was studied by \citet{Schaarsberg2016} before. 

In context of this work, especially the rather small viscosity of the UV-irradiated H$_2$O-CH$_3$OH-NH$_3$-ice is interesting. The according Stokes number is on the order of $St=10^{-2}-1$ and the ice-shell can be expected to act as an effective damper. To put this in perspective, the associated viscosity is comparable to the viscosity of peanut butter, for example \citep{Citerne2001}. The viscosity of the UV-irradiated amorphous water-ice is on the order of $St=10^{-8}-10^{-6}$ and is expected to behave rather solid-like in collisions.

If this effect was present in the PPD, this would probably fundamentally change the dynamics of sticking collisions in certain regions of the disc. CH$_3$OH and NH$_3$ are abundant in PPDs in general \citep{Pontoppidan2014}. \citet{Martin2014} shows that ices consisting of both species desorb at similar temperatures as H$_2$O-ice in a PPD. This is also the reason why NH$_3$ can be considered as a tracer for the H$_2$O-snowline \citep{Zhang2018}. It therefore can be assumed that also mixtures of H$_2$O-CH$_3$OH-NH$_3$-ice can at least theoretically exist within the PPD. As mentioned, while having a smaller viscosity than the bulk-material, collision events of aggregates coated with H$_2$O-CH$_3$OH-NH$_3$-ice might stick at higher collision velocities and overcome the bouncing barrier when irradiated with UV. 

\citet{Tachibana2017} estimate the needed irradiation dose for the liquid-like behaviour of the H$_2$O-CH$_3$OH-NH$_3$-ice to 0.3 photons per molecule. At a distance of 5 au, the time needed to reach the dose can be estimated with the frequency of UV-radiation ($\sim 10^{16}$ Hz) and the UV flux.
The UV-flux is an important quantity as the dose of the UV-photons regulates the liquid-like behaviour of the ice species. Throughout the literature, different UV-fluxes are used which consider different conditions. 
\citet{Walsh2012} describe an UV-flux in a PPD for a typical T Tauri star (viscous parameter $\alpha=0.01$, mass of 0.5 $M_\odot$). The resulting flux depends highly on the height to radius ratio $Z/R$ in the disc. For $Z/R=0$, the mentioned dose will be reached after approximately several hours. For $Z/R=0.15$ the needed time decreases to a value below a millisecond. More pessimistic UV-fluxes are given by \citet{Ciesla2012, Ciesla2010} (with $\alpha=0.001$, 0.1 $M_\odot$). Here, the critical dose is approximately reached after several hours for $Z/R=0.15$ and $10^4$ years and more close to the midplane. The liquid-like behaviour of the ice species is thus more likely to be present at the surface of the PPD. The time for reaching the needed irradiation dose seems none the less to depend highly on the made assumptions of the discs' structure.

It seems reasonable to expect aggregates at the bouncing barrier which might be coated with H$_2$O-CH$_3$OH-NH$_3$-ice behaving liquid-like at least locally in the PPD. In this context, collisions of aggregates coated by this liquid-like layer of H$_2$O-CH$_3$OH-NH$_3$-ice are studied numerically in this work. Collisions of water-ice-coated aggregates with the comparably high viscosity given by \citet{Tachibana2017} are not considered here as they would probably behave much more elastically. The COR is calculated for different sets of parameters as well as the change of the sticking velocity compared to "naked" aggregates. 

\section{The Collision Model}

\subsection{The Particle-Particle Interaction}

The collision model in this work assumes an impact between two identical, spherical particles by which collisions of aggregates in a PPD are modelled. The basis of this model is the contact model introduced by \citet{Thornton1998}. \citet{Thornton1998} consider sticking forces, elastic forces and plastic forces and give an equation for the coefficient of restitution (COR). The COR is described by the impact velocity $v_\text{i}$ of two colliding bodies and their velocity after collision $v_\text{o}$ and is given by the ratio $\epsilon(v_\text{i})\equiv v_\text{o}/v_\text{i}$. \citet{Thornton1998} calculate the COR depending only on the impact velocity $v_\text{i}$, on the sticking velocity $v_\text{s}$ and on the velocity for the onset of plastic deformation $v_\text{pl}$ within the collision bodies. The full expression is calculated in the original work as a continuous, piecewise function (sticking regime, elastic regime and plastic regime) and will be quoted as $\epsilon_\text{TN}(v_\text{i},v_\text{s},v_\text{pl})$ in further. 
Additionally to the collision of the two spherical aggregates, a liquid-like shell covering both aggregates is considered in this work. Before and after the collision described by \citet{Thornton1998}, this shell has to be crossed through by the aggregates which acts as an additional damping. For the conditions covered here the Reynolds number is $\text{Re}\ll 1$. For this reason, a drag force given by Stokes' law is assumed. The equation of motion is then given by
\begin{equation}
\frac{d^2}{dx^2}x(t)+\frac{6\pi\eta R}{m}\frac{d}{dx}x(t)=0,
\label{stokeslaw}
\end{equation}
where $x(t)$ is the distance between the aggregates and $\eta$ the dynamic viscosity. $R$ is the effective radius for the drag force which is represented by the intersection of both ice-shells as an approximation. The assumption behind this is that the force transmission has to be limited to this intersection in the best case. For the simulation, $\eta=5\cdot10^2$ Pa$\cdot$s \citep{Tachibana2017} and a density of 2890 kg/m$^3$ (basalt) for the aggregates are assumed. The effective radius $R$ depends on the current relative position of the aggregates. The value is calculated by the mean value theorem for integration and changes with the thickness of the shell. These assumptions are made to represent basaltic aggregates at the bouncing barrier coated by an ice-shell of H$_2$O-CH$_3$OH-NH$_3$ at 5-10 au in the PPD. 

A solution to eq.\eqref{stokeslaw} is given by
\begin{equation}
    x(v_\text{i},t)=\frac{mv_\text{i} }{6\pi\eta R}\left(1- \exp\left(-\frac{6\pi\eta R}{m}t\right)\right).
\end{equation}
The relative velocity between both aggregates decreases exponentially in time. With given initial conditions, the collisional time $t_\text{c,b}$ during the motion through the ice-shell till the contact of both inner aggregates can be calculated from this solution. The condition is given by 
\begin{equation}
  x(v_\text{i},t_\text{c,b}) \stackrel{!}{=} 2d_\text{ice},  
\end{equation}
where $d_\text{ice}$ is the layer thickness of the ice shell. This finally allows to determine the velocity after the damping caused by the ice-shell on the inward motion $v(v_\text{i},t_\text{c,b})$. 

\begin{figure}
    \centering
	\includegraphics[width=0.8\columnwidth]{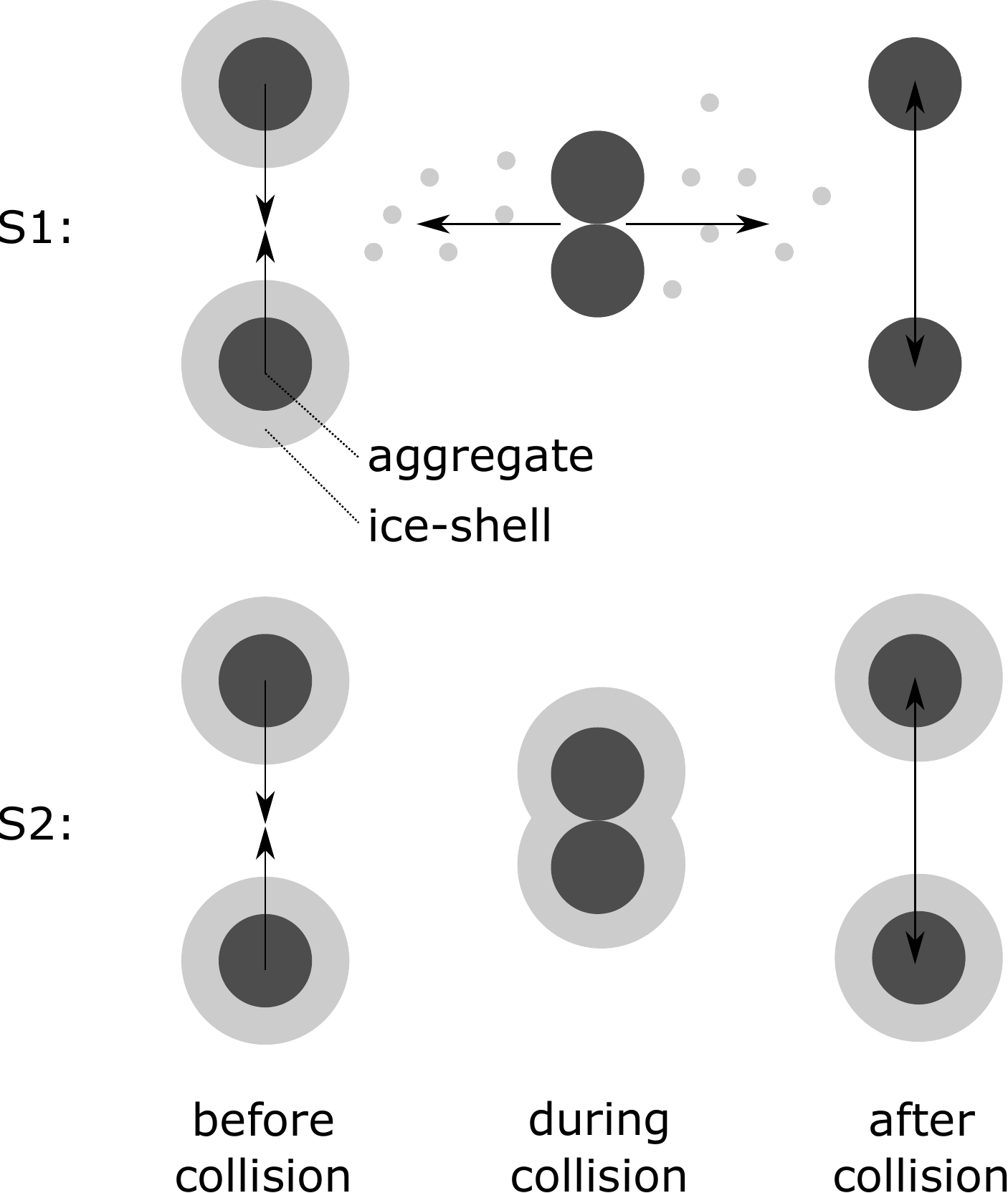}
	\caption{\label{fig.cases}Two limiting cases covered by the collision model in this work. In the scenario S1 the ice shell is fragmented after the collision of the aggregates. The ice-shell contributed to damping only before the collision takes place. In the scenario S2 the ice-shell remains intact and damps the aggregates before and after the collision.}
\end{figure}

The effective COR can be deduced from the presented calculation. In this work, two limiting cases are considered as depicted in fig. \ref{fig.cases}.
In the first scenario S1, the aggregates are damped by the ice shell only before the collision of the inner aggregates takes place. This scenario assumes that the ice shell is fragmented completely after the impact as kinetic energy might be transferred to fragmentation energy of the shell partially. Here, the effective COR for the whole collision process can be calculated for constant $v_\text{s}$ and $v_\text{pl}$ as
\begin{equation}
    \epsilon_\text{S1}(v_\text{i},t_\text{c,b})= \frac{v(v_\text{i},t_\text{c,b})}{v_\text{i}}\epsilon_\text{TN}(v(v_\text{i},t_\text{c,b})).
    \label{cors1}
\end{equation}
This expression for the COR is a product of the effective COR for the movement through the ice-shell and additionally the COR for the aggregates without any ice-shell described by \citet{Thornton1998}. 

In the second scenario S2, the aggregates are additionally damped by the ice-shell while moving outwards. In this case the shell is assumed to be "sticky enough" to avoid any fragmentation. The time of the aggregates' motion through the ice shell is given by $t_\text{c,a}$ and the velocity after the collision process by $v(\epsilon_\text{S1}v_\text{i},t_\text{c,a})< v(v_\text{i},t_\text{c,b})$. The COR in this scenario has then the expression
\begin{equation}
    \epsilon_\text{S2}(v_\text{i},t_\text{c,b},t_\text{c,a}) 
    =\frac{v(\epsilon_\text{S1}(v_\text{i},t_\text{c,b})v_\text{i},t_\text{c,a})}{\epsilon_\text{S1}(v_\text{i},t_\text{c,b})v_\text{i}}\epsilon_\text{S1}(v_\text{i},t_\text{c,b}),
\label{cors2}
\end{equation}
and is smaller than $\epsilon_\text{S1}(v_\text{i},v_\text{s},v_\text{pl},t_\text{c,b})$ for every impact velocity $v_\text{i}$. It should be noted that in any real collision of aggregates the ice-mantle might partially fragment. Thus, real collisions take place somewhere between S1 and S2. The model does not cover any dissipation through sound waves on the surface of the aggregates due to the impact. 

The COR is calculated for a sticking velocity of $v_\text{s}=0.005$ m/s. This represents approximately the sticking velocity of dusty aggregates at the bouncing barrier \citep{Guettler2010}. Furthermore, the two cases $v_\text{pl}\in \{0.5 \text{m/s}, 5 \text{m/s}\}$ are analysed. In comparison to this, the fragmentation velocity of loosely bound aggregates consisting of $\sim 100 \mu$m-large sub-aggregates with a maximum tensile strength of 100 Pa (which e.g. were analysed by \citet{Musiolik2017} as an analog for cometary dust) is on the order of $0.5-1$ m/s. For aggregates at the bouncing barrier, \citet{Guettler2010} give a fragmentation velocity on the order of 1 m/s. The second case of $v_\text{pl}=5$ m/s describes more rigid aggregates to illustrate the dependency of the COR on varying $v_\text{pl}$.

Finally, the sticking velocity of the ice-shell mantled aggregates can be estimated numerically. The condition for the sticking velocity is
\begin{equation}
    v_\text{o}=v_\text{i}\cdot\epsilon_{\text{S1}\oplus\text{S2}}\stackrel{!}{=}0.
    \label{stickingcondition}
\end{equation}
This value is calculated for different thicknesses of the ice-shell to a precision of $10^{-8}$ m/s. With increasing thickness of the ice-shell the sticking velocity has to increase as well by definition.

\subsection{Numerical Growth Model}
The calculated COR and sticking velocity can be used to simulate the collisions between a large number of aggregates with a defined ice-shell layer thickness numerically. This is done by the Monte Carlo method. Here, only the first scenario S1 is considered as limiting case. This can be used as a "worst case" prediction for the growth of larger clusters inside a PPD. 

Initially, all aggregates have normally distributed velocities. The mean relative velocity between two random aggregates is $\sim 0.35$ m/s, which is in line with relative velocities of $10^{-3}$ m-sized aggregates inside a PPD as e.g. calculated by \citet{Windmark2012}. A collision event occurs between two randomly chosen aggregates or clusters (consisting of several aggregates) and has three possible outcomes: 
\begin{itemize}
    \item[1)] If the relative velocity is below the sticking velocity of the aggregates/clusters covered by the ice-shell, the aggreates/clusters stick to each other. The sticking velocity is calculated with eq.\eqref{stickingcondition}.
    \item[2)] If the relative velocity is larger than the sticking velocity, the aggregates/clusters bounce off as described by eq.\eqref{cors1} and recouple to the gas in the PPD. A  normally distributed velocity (analog to the initial conditions) is assigned to both.
    \item[3)] If the relative velocity is larger than the fragmentation velocity, clusters are critically fragmented into individual monomers. This is a simplification in the computation similar as in the fragmentation model developed by \citet{Dullemond2005, Zsom2010phd}. Inside a PPD clusters would probably fragment only partially. None the less, the simplified fragmentation model can be considered as representative for a limiting case. The disintegrated monomers from the cluster recouple to the gas in the PPD. A normally distributed velocity is assigned to them. Individual monomers cannot fragment. This is not critical, because the growth of aggregates up to $10^{-3}$ m in a PPD is not considered as problematic. The fragmentation velocity is chosen to 1 m/s based on the values given by \citet{Guettler2010}.
\end{itemize}

The simulation performed in this work comprises $10^5$ collisions of $10^4$ aggregates with a size of $10^{-3}$m each (no mass distribution) and is repeated 20 times for several thicknesses of the ice-shell which covers the aggregates. The cluster-size-distribution reaches an equilibrium state and can be deduced from the simulation. Another important quantity which is calculated as well is the largest cluster within the equilibrium state. This is especially important to estimate the impact on growth of larger clusters inside a PPD afterwards.

\section{Results and Interpretation}
\subsection{Coefficient of Restitution and Sticking Velocity}
\begin{figure*}
    \centering
	\includegraphics[width=0.99\columnwidth,valign=t]{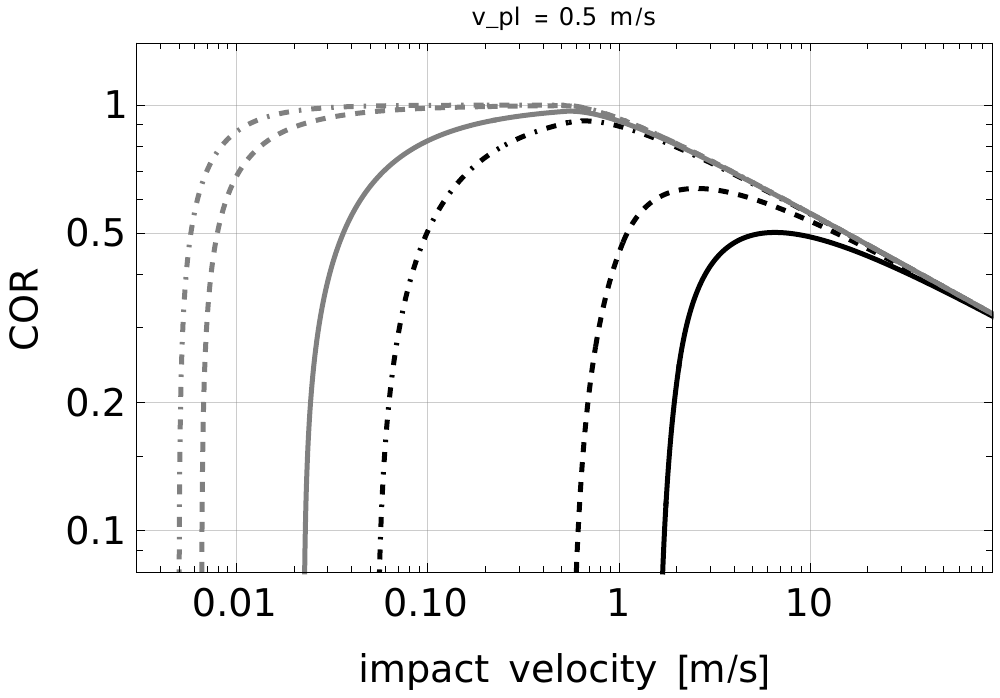}
	\includegraphics[width=0.99\columnwidth,valign=t]{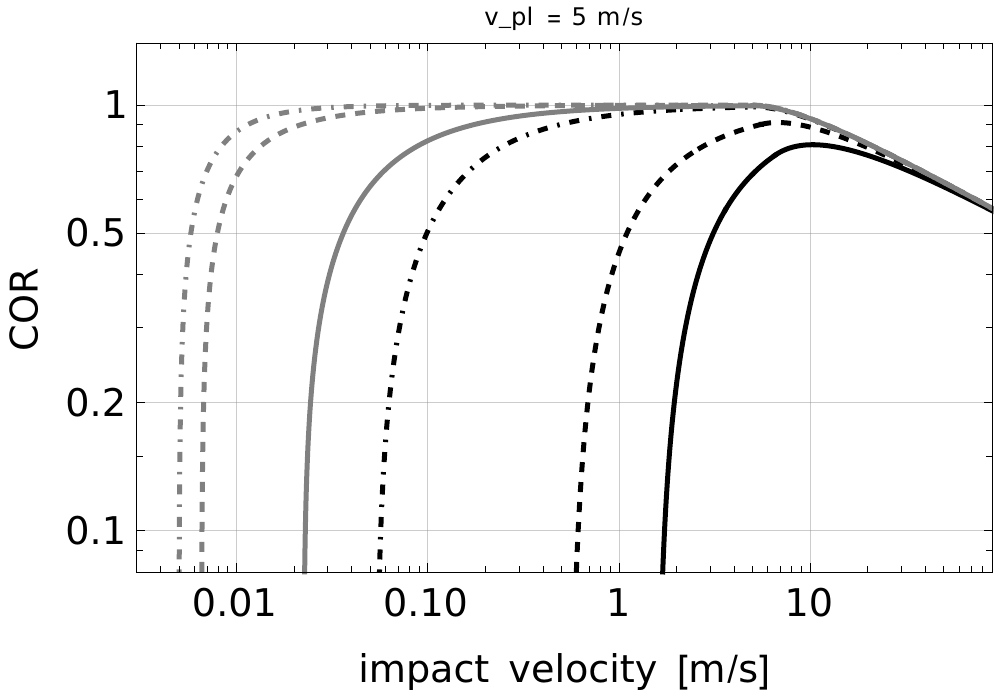}\newline
	\includegraphics[width=0.99\columnwidth,valign=t]{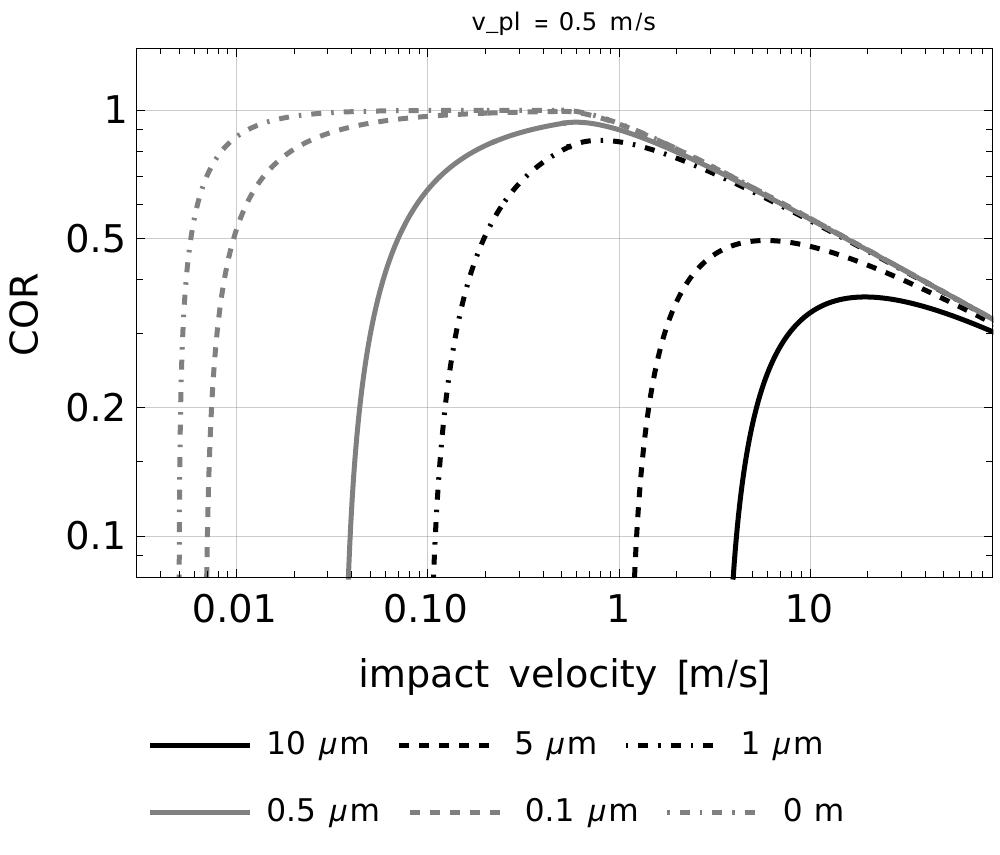}
	\includegraphics[width=0.99\columnwidth,valign=t]{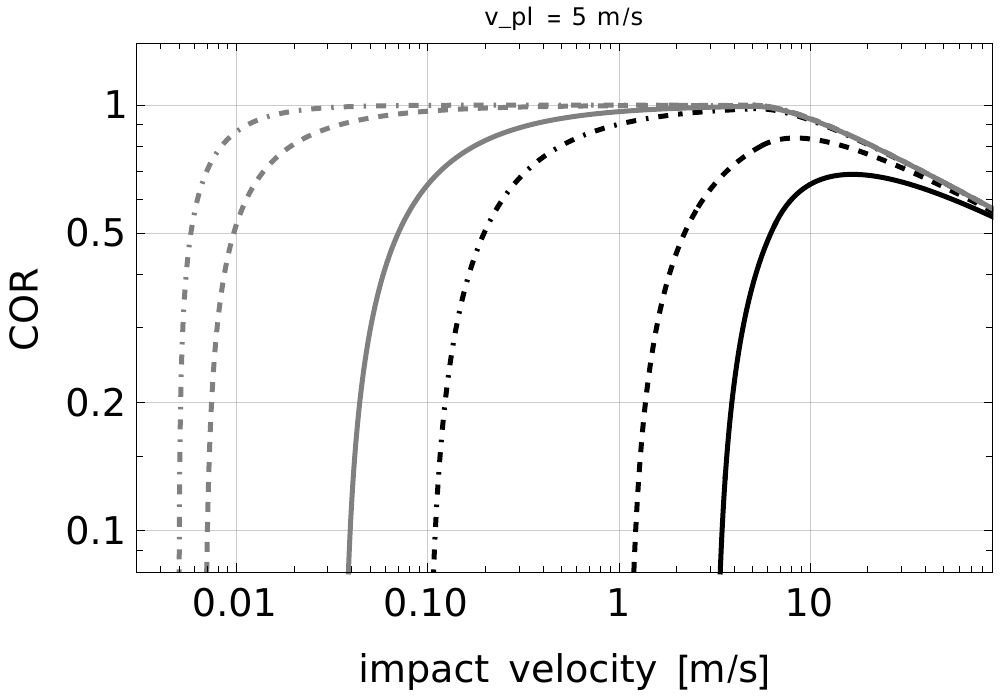}
	\caption{\label{fig.cors}The COR depending on the impact velocity for the scenario S1 (top row) and for scenario S2 (bottom row) and $v_\text{pl}=0.5$ m/s (left column) as well as $v_\text{pl}=5$ m/s (right column). The different functions belong to different thicknesses of the ice-shell as described in the legend.}
\end{figure*}

Fig. \ref{fig.cors} presents the COR in the case of scenario S1 (top row) for various thicknesses of the ice-shell $d_\text{ice}$ ranging from $0.1 \mu$m to $10 \mu$m and for $v_\text{pl}=0.5$ m/s as well as for $v_\text{pl}=5$ m/s. 

The ice-shell is fully fragmented due to the impact of the collision in this case. Three important findings can be deduced from the calculations at this point. 
\begin{itemize}
    \item[1)] The sticking velocity increases with a thicker ice-shell layer as expected. This increase gets significant for a layer thickness on the order of $1 \mu$m and would be sufficient to raise the sticking velocity over a factor of 10 compared to aggregates without any ice-shell. Aggregates with a layer thickness on the order $0.1 \mu$m behave similar to aggregates without any ice-shell. This might not surprise as an ice-shell of $0.1 \mu$m is equivalent to not more than 100 mono-layers of ice molecules on the aggregates.
    \item[2)] For the COR, the relation $\epsilon_\text{S1}(d_\text{ice})\leq\epsilon_\text{TN}$ applies for all impact velocities. For large velocities, $\epsilon_\text{S1}(d_\text{ice})$ converges towards the COR of aggregates without any ice-shell $\epsilon_\text{TN}$. 
    \item[3)] An increase in $v_\text{pl}$ extends the elastic collision behaviour to greater impact velocities. For thicknesses of the ice layer where the difference between the sticking velocity and $v_\text{pl}$ are small the COR hence reaches slightly higher maximum values. At the same time, the sticking velocities remain unchanged.
\end{itemize}

In the second scenario S2 the ice-shell remains intact after the collision. The calculated COR (using eq.\eqref{cors2} in this case) is presented in fig. \ref{fig.cors} (bottom row) for $v_\text{pl}=0.5$ m/s and for $v_\text{pl}=5$ m/s. The following findings can be derived from the COR-plots in this case. 
\begin{itemize}
    \item[1)] The COR in scenario S2 is smaller than the COR in scenario S1 for all impact velocities and a constant $v_\text{pl}$. This behaviour is expected because the aggregates have to cross through an additional ice-shell layer. 
    \item[2)] Due to the additional damping, the sticking velocity is higher than in scenario S1 for every (non-zero) layer-thickness. For example, a layer thickness of $1 \mu$m is responsible for an increase in the sticking velocity by a factor of approximately 2 compared to scenario S1. This seems reasonable as the thickness of the layer to be crossed raises by a factor of 2 as well in this case.
    \item[3)] Increasing $v_\text{pl}$ results in an wider elastic regime comparable to scenario S1.
\end{itemize}

\begin{figure}
    \centering
	\includegraphics[width=0.99\columnwidth]{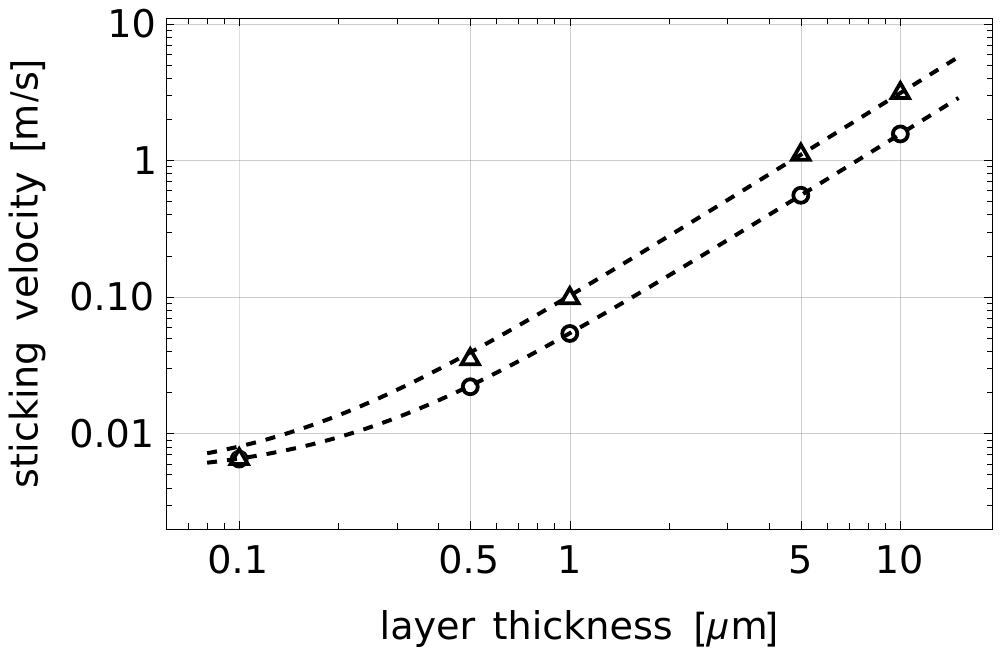}
	\caption{\label{fig.vstick}Sticking velocity for $10^{-3}$ m large dust aggregates depending on the thickness of the ice-shell layer. The bottom line (circles) shows S1 where the ice-shell fragments after the collision. The top line (triangles) shows S2 where the ice-shell remains intact after collision. Note, that the necessary condition for the validity of these values is $v_\text{s}(d_\text{ice})\leq v_\text{pl}(0)$ as discussed in detail below. This might not apply for layer thicknesses around $10 \mu$m m on dust aggregates any more.}
\end{figure}

An important quantity for the further analysis is the sticking velocity $v_\text{s}$. The value can be calculated from the COR using eq.\eqref{stickingcondition}. The dependency between $v_\text{s}$ and the thickness of the ice-shell layer $d_\text{ice}$ is shown in fig. \ref{fig.vstick}.
Both of the extreme cases S1 and S2 are plotted in this figure. The sticking velocity in scenario S2 is higher than in S1 for all simulated points. This difference increases with the thickness of the ice-shell (absolute value). For a very narrow ice-shell layer on the scale of $0.1 \mu$m the sticking velocity converges towards the initial $v_\text{s}(0)=0.005$ m/s (aggregates without any ice-shell). The fits shown in fig. \ref{fig.vstick} are given by the function
\begin{equation}
    v_\text{s}(d_\text{ice})=v_\text{s,0}+\alpha d_\text{ice}^\beta
    \label{vdeq}
\end{equation}
with the layer thickness $d_\text{ice}$ and the fitting parameters $\alpha,\beta$. In both scenarios S1 and S2, the exponent in the fit is given to $\{\beta\}=1.5\pm0.05$. For S1, the other parameter is given to $\{\alpha_\text{S1}\}=0.049$. For S2, the value changes to $\{\alpha_\text{S2}\}=0.097$. 

For a macroscopic ice-layer thus $v_\text{s}\propto d_\text{ice}^{3/2}$ can probably be assumed. The dissipated energy in the collision would then be $E_\text{diss}\propto v^2_\text{s}\propto d_\text{ice}^{3}$ according to the sticking model by \citet{Thornton1998}. The reason for this dissipation is a combination of two effects. The first effect is the transfer of kinetic energy of the aggregates to the fluid (Stokes' law). The second effect is the energy dissipation due to the surface and plastic forces of the aggregates' inner dust-cores.

A more realistic model could also comprise other effects like the transfer of heat energy, the propagation of sound waves or the displacement of the shell as well. This however is not within the scope of this work and has to be analysed in more detail in future. 

Note, that the real sticking behaviour is expected to take place somewhere between both extremes S1 and S2 and depends on the material of the collision bodies. In general, sticking collisions with small $v_\text{i}$ will tend to behave like in scenario S2 whereas sticking collisions with high $v_\text{i}$ will more likely behave as in scenario S1. This is caused by the fact that more collision energy is transferred into restructure/fragmentation energy at high impact velocities. This effect can be covered by eq.\eqref{vdeq} by considering the fitting parameters to depend on $v_\text{s}$. 

At this point, also another important effect has to be considered. Aggregates with a small $v_\text{pl}$ will begin to fragment earlier by increasing the impact velocity in collisions. For any case in which $v_\text{s}$ for the aggregates with an ice-shell is greater than $v_\text{pl}$ for aggregates without an ice-shell a further growth in sticking collisions might not be possible. Instead, the aggregates would more likely tend to disintegrate before sticking to each other. Note that $v_\text{pl}$ is an indicator which describes the transition between collisions in which fragmentation is not allowed and in which it might occur. The real fragmentation velocity (defined often as the velocity at which more than half of the aggregate disintegrates) has to be none the less larger than $v_\text{pl}$. In other words, the necessary condition for the validity of fig. \ref{fig.vstick} is $v_\text{s}(d_\text{ice})\leq v_\text{pl}(0)$. For this reason, especially the sticking velocity for the layer thickness around and above $10 \mu$m has to be treated with caution because it might not result in effective growth of the colliding aggregates.

These results predict an important implication for the collisions of aggregates with an ice-shell as analysed inside a PPD. \citet{Windmark2012} calculate the relative collision velocities for aggregates in a PPD depending on their size. For aggregates on the $10^{-3}$ m scale, the collision velocities are approximately $0.35$ m/s. Compared with the curve from fig. \ref{fig.vstick} this means that an ice-shell layer with a thickness of $2-3 \mu$m would be sufficient to expect sticking collisions. At these collision velocities, fragmentation should be negligible to a large extent. 

But how does this relate to the abundances of the ice species inside a PPD? Can a $2-3 \mu$m ice-shell be expected here? \citet{Boogert2008,Bottinelli2010} estimate the relative abundances for low luminosity YSOs from observations and give a CH$_3$OH to H$_2$O ratio between 0.01-0.03. The ratio of NH$_3$ to H$_2$O is given to 0.03-0.08 by the authors. \citet{Pontoppidan2014} derive the abundance ratio between silicates and H$_2$O for YSOs which is approximately 1.5. If one assumes that the abundance of H$_2$O-CH$_3$OH-NH$_3$ is limited to the smallest possible abundance of CH$_3$OH this would result in a H$_2$O-CH$_3$OH-NH$_3$ (molecular ratio of 5:1:1 as in the work of \citet{Tachibana2017}) to silicate ratio of $\chi\approx0.01/(5\cdot 1.5)\approx0.00133$. For a size of $r_\text{core}=10^{-3}$ m of the silicate-core the size of the ice-shell can then be estimated to
\begin{equation}
    r_\text{shell}=\chi^\frac{1}{3}r_\text{core}\approx 1.1\cdot10^{-4} \text{m},
\end{equation}
which is at least an order of magnitude more than the needed $2-3 \mu$m ice-shell size. 
This means that sticking collisions at the bouncing barrier described by the mechanism considered in this work could be realistic, indeed. In the further, the impact of this is studied for an ensemble of $10^5$ colliding aggregates in a simulation.

\subsection{Cluster-Growth in a PPD}
The growth of clusters inside a PPD is modelled by the algorithm presented before. The calculation shown below is only done for scenario S1 as the "worst case" for what can be expected in a PPD. 

\begin{figure*}
    \centering
	\includegraphics[width=0.99\columnwidth,valign=t]{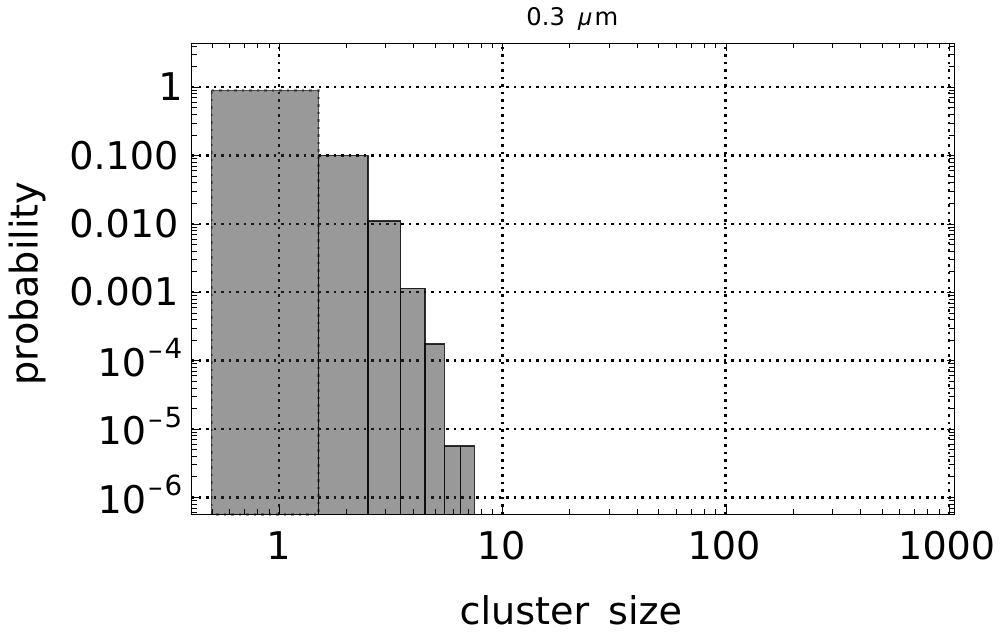}
	\includegraphics[width=0.99\columnwidth,valign=t]{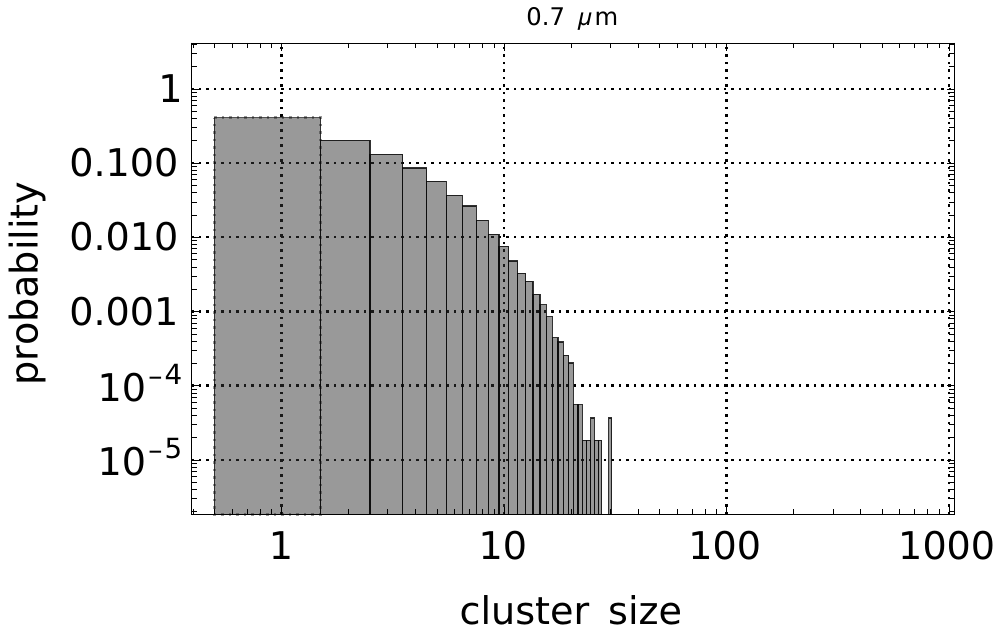}\newline
	\includegraphics[width=0.99\columnwidth,valign=t]{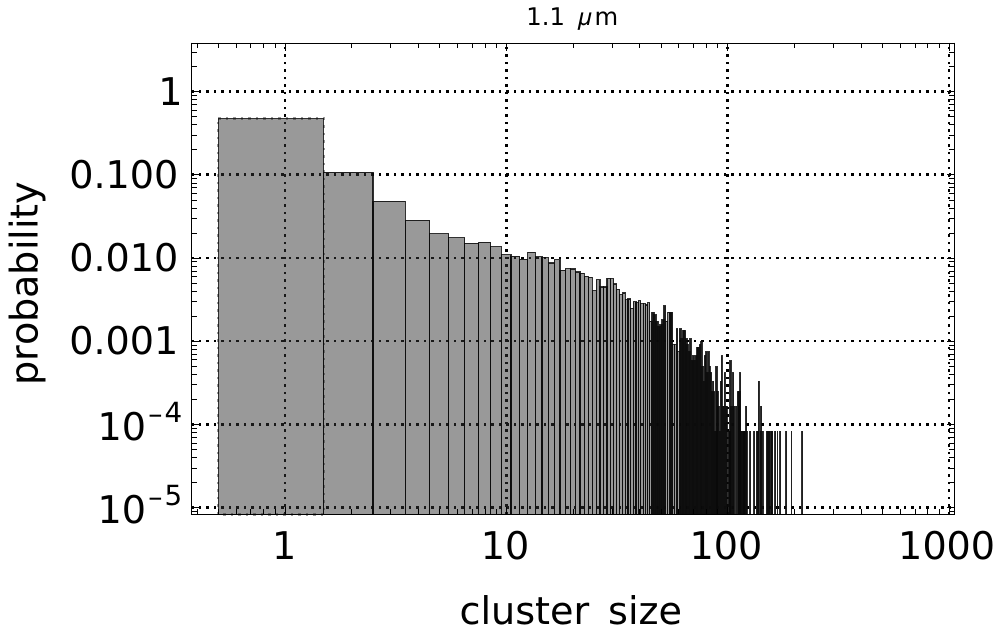}
	\includegraphics[width=0.99\columnwidth,valign=t]{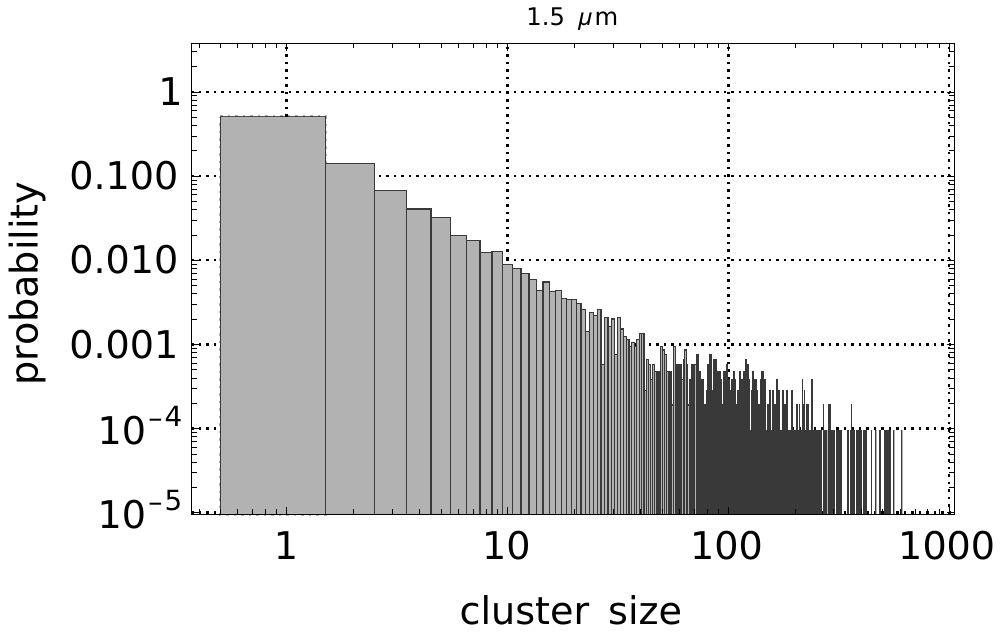}
	\caption{\label{fig.hist}Probability to find a cluster with a certain cluster-size on a log-log-scale. The different figures indicate the transition of aggregates coated with different thickness of the ice-shell reaching from $0.3 \mu$m (top, left) to $1.5 \mu$m (bottom, right). The data represents the equilibrium state. This means that further collisions between aggregates/clusters do not affect the shape of the distribution.}
\end{figure*}

The simulation shows that in the equilibrium state, the distribution of the cluster-size depends on the thickness of the ice-shell. This dependency is presented in fig.\ref{fig.hist}.
For a small shell-thickness of $0.3 \mu$m, most of the aggregates have a size of 1-3 aggregates. With increasing shell-thickness the distribution becomes "flatter" and the clusters grow on average. Interestingly, the size distribution is nearly given by a power law for small cluster sizes (relative to the largest fragment) which is in accordance to fragmentation experiments as e.g. described by \citet{Teiser2009,Guettler2010,Musiolik2016a}. This is clearly visible especially for the $1.5 \mu$m ice-shell. None the less, for mid to large cluster sizes this dependency seems to be violated. 

\begin{figure}
    \centering
	\includegraphics[width=0.99\columnwidth]{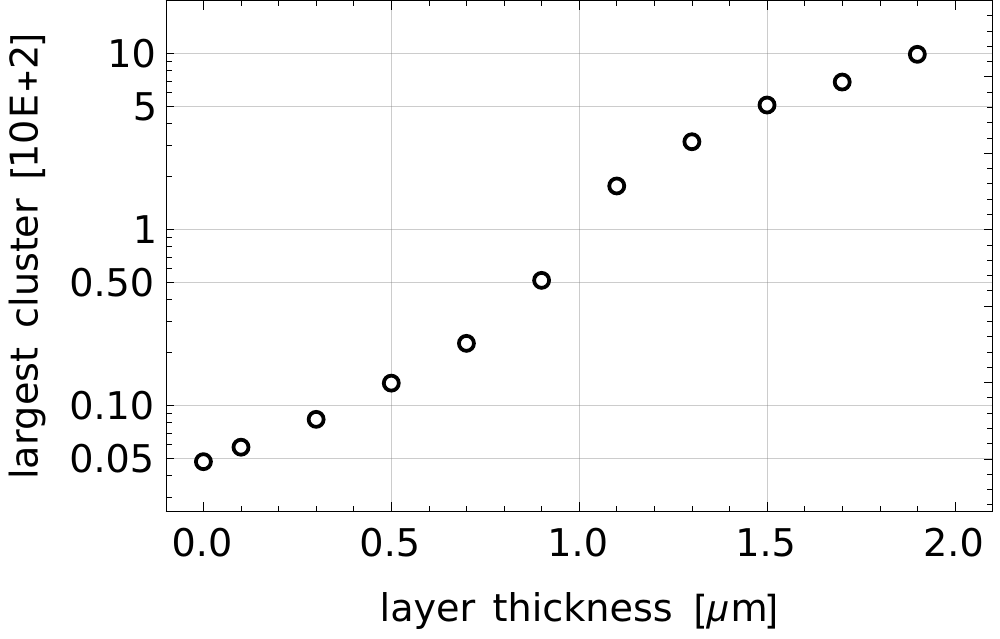}
	\caption{\label{fig.lafrag}The largest cluster in the simulation space in dependence on the layer thickness of the ice-shell on aggregates.}
\end{figure}

For a given thickness of the ice-shell there is a defined cut-off for the largest possible cluster to grow. This largest fragment can be deduced from the simulation as well. The size of the largest fragment increases with the shell-thickness which is not surprising. Fig. \ref{fig.lafrag} depicts this behaviour.
A simple dependency e.g. a power-law or a exponential law cannot be deduced from the data. However, an interesting size-transition of the largest cluster is observable around a thickness of $1 \mu$m in the simulation. For this shell-thickness, the largest cluster contains approximately 100 aggregates (with a size of $10^{-3}$ m each). If these aggregates were in a close-packing forming a spherical cluster, the cluster itself had a size of $0.89$ cm. This of course would require an equal distribution of the impact events in all radial directions in a PPD. With a porosity of $0.5-0.2$ the cluster size changes to $1.1-1.5$ cm. For a shell-thickness of $1.9 \mu$m, the cluster would already contain 1000 aggregates. Similarly in this case, the size of the cluster was $1.9$ cm. With a porosity of $0.5-0.2$ this value increases to $2.4-3.3$ cm. 

The size of the largest cluster saturates with increasing layer thickness. This behaviour is caused by the fragmentation barrier. Larger clusters do not collide at higher impact velocities in the simulation but they yet disintegrate critically each time in a fragmentation event. Thus, the growth of even larger clusters freezes at this fragmentation barrier so any additional thickness of the layer does not have any supportive effect for the growth of the clusters.

Within the PPD such behaviour would allow the growth transition towards cm-sized pebbles. Due to their comparably large mass, pebbles are supposed to drift radially inwards and vertically towards the mid-plane. This effect is not covered by the simulation in this work. Due to the still rather large time-scales for the drift this might however be negligible at least for $\sim$ cm-sized pebbles \citep{Hu2014}.

\section{Conclusion}
The collision growth of dust aggregates in a PPD is interrupted at the bouncing-barrier at a size of $10^{-3}$m. The aggregates can be mantled by different types of ice which changes the collision behaviour and might shift the bouncing barrier to larger values. In this work, a shell of H$_2$O-CH$_3$OH-NH$_3$-ice on the aggregates is examined. 

This ice-species behaves liquid-like under certain conditions like UV-irradiation and a low temperature which are comparable to the environment in a PPD between 5-10 au as described by the MMSN. In collision events, aggregates with this ice-shell would be damped due to Stokes' drag. The effective COR indicates that aggregates at the bouncing barrier could then stick to each other within a PPD on average once having an ice-shell with a thickness of $2-3 \mu$m. 

The numerical simulation shows that an ice-shell with the thickness of a micron could already result in the growth of a cm-sized cluster. This could be an enhancing mechanism for the further steps in the planet forming process in a PPD.  

Similar mechanisms might not only be possible for UV-irradiated H$_2$O-CH$_3$OH-NH$_3$-ice. If photodissociation is resposible for this effect, $\gamma$-radiation might have an analogical impact for other ice species like H$_2$O which was not considered here. This would increase the density of aggregates with large sticking velocities in a PPD even more.

The shown results are theoretical and computational considerations. Further experimental work on collisions of aggregates can show if this effect is reproducible in the laboratory as well and therefore verify its potential to boost the growth of pebble-sized aggregates in the PPD.

\section*{Acknowledgements}
The author appreciates the constructive review from Shogo Tachibana. The author did not receive any specific funding for this work.

\section*{Data Availability}
The data that support the findings of this work are available from the corresponding author upon reasonable request








\bsp	
\label{lastpage}
\end{document}